\documentclass[10pt,twocolumn,letterpaper,breaklinks=true,bookmarks=false]{article}

\usepackage[utf8]{inputenc}
\usepackage{iccv}
\usepackage{soul}
\soulregister\cite7
\soulregister\ref7
\soulregister\pageref7
\soulregister\textquotesingle7
\usepackage{times}
\usepackage{epsfig}
\usepackage{graphicx}
\usepackage{amsmath}
\usepackage{amssymb}
\usepackage{subfig}
\usepackage{url}
\usepackage{booktabs}
\usepackage{pythonhighlight}
\usepackage{hyperref}
\usepackage[square,numbers]{natbib}

\iccvfinalcopy 

\ificcvfinal\fi

\begin{document}
\let\WriteBookmarks\relax
\def\floatpagepagefraction{1}
\def\textpagefraction{.001}

\title{3D Convolutional Neural Networks for Stalled Brain Capillary Detection}

\author{Roman Solovyev\\
Institute for Design Problems in Microelectronics of Russian Academy of Sciences\\
3, Sovetskaya Street, Moscow 124365, Russian Federation\\
{\tt\small roman.solovyev.zf@gmail.com}
\and
Alexandr A. Kalinin\\
Shenzhen Research Institute of Big Data\\
Shenzhen 518172, Guangdong, China\\
Department of Computational Medicine and Bioinformatics\\ 
University of Michigan, Ann Arbor, MI 48109, USA
\and
Tatiana Gabruseva\\
Cork University Hospital\\
Cork, Ireland\\
{\tt\small tatigabru@gmail.com}
}

\maketitle

\begin{abstract}
Adequate blood supply is critical for normal brain function.
Brain vasculature dysfunctions, including stalled blood flow in cerebral capillaries, are associated with cognitive decline and pathogenesis in Alzheimer’s disease.
Recent advances in imaging technology enabled generation of high-quality 3D images that can be used to visualize stalled blood vessels.
However, localization of stalled vessels in 3D images is often required as the first step for downstream analysis.
When performed manually, this process is tedious, time-consuming, and error-prone.
Here, we describe a deep learning-based approach for automatic detection of stalled capillaries in brain images based on 3D convolutional neural networks.
Our approach includes custom 3D data augmentations and a weights transfer method that re-uses weights from 2D models pre-trained on natural images for initialization of 3D networks.
We used an ensemble of several 3D models to produce the winning submission to the "Clog Loss: Advance Alzheimer’s Research with Stall Catchers" machine learning competition that challenged the participants with classifying blood vessels in 3D image stacks as stalled or flowing.
In this setting, our approach outperformed other methods and demonstrated state-of-the-art results, achieving 85\% Matthews correlation coefficient, 85\% sensitivity, and 99.3\% specificity.
The source code for our solution is publicly available.
\end{abstract}

\section{Introduction}
\label{intro}
Healthy brain vasculature is crucial for maintaining cerebral blood flow (CBF) that supplies neurons with oxygen, energy metabolites, and nutrients. It removes carbon dioxide and other potentially toxic metabolic waste products from the brain and into the systemic circulation for clearance \cite{zlokovic2011neurovascular,kisler2017cerebral}.
With limited energy reserve, brain functions stop within seconds if CBF stops, and irreversible damage to neurons occurs within minutes \cite{moskowitz2010science,zlokovic2011neurovascular}.
It has been widely acknowledged that disruption of normal CBF is an early and persistent symptom in developing Alzheimer’s disease (AD) and other neurodegenerative disorders \cite{moskowitz2010science,zlokovic2011neurovascular,kisler2017cerebral,tang2019model,sweeney2018role,bracko2021causes}.
However, underlying mechanisms for cerebral blood flow reduction in Alzheimer’s disease are not well understood.

Recent advances in imaging technologies, such as multiphoton microscopy, enabled 3D visualization of individual capillaries in living tissue up to about one millimeter in thickness \cite{hernandez2019neutrophil,haft2019deep}.
The ability to perform this type of imaging in mouse models of AD offers the opportunity to elucidate the mechanistic links between CBF reductions and AD pathology \cite{hernandez2019neutrophil,bracko2021causes}.
Typically, the analysis of such images starts with identifying capillaries and then manually labeling each capillary segment as either flowing or stalled based on the motion of blood cells during the entire time each capillary is visible in the three-dimensional image stack \cite{hernandez2019neutrophil,bracko2020increasing}.
However, the process of manual annotation is very tedious and time-consuming, which limits the ability to investigate the vital link between capillary function and AD.

One increasingly popular approach to data annotation that aims to speed up data annotation is crowdsourcing~\cite{michelucci2016power}, which combines the efforts of many individuals tasked with annotating a small part of a larger data set.
This approach creates an opportunity to use the crowdsourced annotations for training deep learning models, which rely on the availability of large labeled datasets~\cite{vaughan2018making}.
Specifically, the growing availability of annotated data has recently enabled efficient applications of deep learning to the analysis of biomedical images~\cite{litjens2017survey}, demonstrating state-of-the-art results in object detection, segmentation and classification~\cite{rakhlin2019breast,solovyev2020bayesian,kalinin2020medical,rsna}.
In multiphoton microscopic images, deep learning have been recently used for segmenting blood vessels \cite{haft2019deep,haft2020topological} and segmenting and classifying cancer cells \cite{lin2019automated,cai2020dense,huttunen2020multiphoton}.

A citizen science project named Stall Catchers (formerly "EyesOnALZ") aims to speed up the identification of stalls in research data using crowdsourcing.
To prepare data for annotation by volunteers in Stall Catchers, raw image stacks of the vasculature are first masked to remove the larger surface vessels and low signal-to-noise regions, and preprocessed to normalize the image intensity distribution and spatial resolution and to remove motion artifacts \cite{bracko2020high}.
After that, capillary regions that need to be annotated are identified using a deep convolutional neural network (CNN) trained to segment cortical blood vessels \cite{falkenhain2020pilot,bracko2020high}.
Finally, the Stall Catchers platform educates and challenges participants with labeling images of individual capillaries as \textit{flowing} or \textit{stalled}.
The resulting labeled data have already enabled new scientific studies \cite{falkenhain2020pilot,bracko2020high}, but they also created an opportunity to further improve the data labeling process itself by the application of deep learning.

Here, we propose an approach for the automatic detection of stalled capillaries in 3D brain images from Stall Catchers data using deep convolutional neural networks.
To address class imbalance, we also introduce several 3D image augmentation transformations and demonstrate how weights from pre-trained 2D models can be re-used in 3D architectures.
Our solution placed $1^{st}$ in the "Clog Loss: Advance Alzheimer’s Research with Stall Catchers" machine learning competition organized by Stall Catchers and DrivenData \cite{lipstein2020winners}.
Our approach was recognized as the most advanced in both performance and sophistication among proposed solutions by the competition organizers, who estimated that our model could automatically analyze about 50\% of all existing Stall Catchers data, effectively doubling the analysis speed and helping to uncover insights towards connections between stalled blood flow and AD \cite{ramanauskaite2020stalls}.

\section{Methods}
\subsection{Problem statement and data description}
The dataset used for the "Clog Loss: Advance Alzheimer’s Research with Stall Catchers" competition was provided by Stall Catchers, a citizen science effort created by the Human Computation Institute~\cite{humancomputation}.
The objective of the challenge was to perform binary classification of the outlined blood vessel segments as either \textit{flowing}, if blood is moving through the vessel, or \textit{stalled}, if the vessel has no blood flow.

Each sample in the dataset contained a set of images (3D image volume) taken from brain tissue of an alive mouse, showing blood vessels and the blood flow through them. Images were taken via multiphoton microscopy and had $\approx0.5$ $\mu$m lateral and $\approx1.5$ $\mu$m axial spatial resolution. The resolution could have been degraded by motion artifacts due to the animal's heartbeat and respiration.

\begin{figure*}[htb]
 \includegraphics[width=0.99\linewidth]{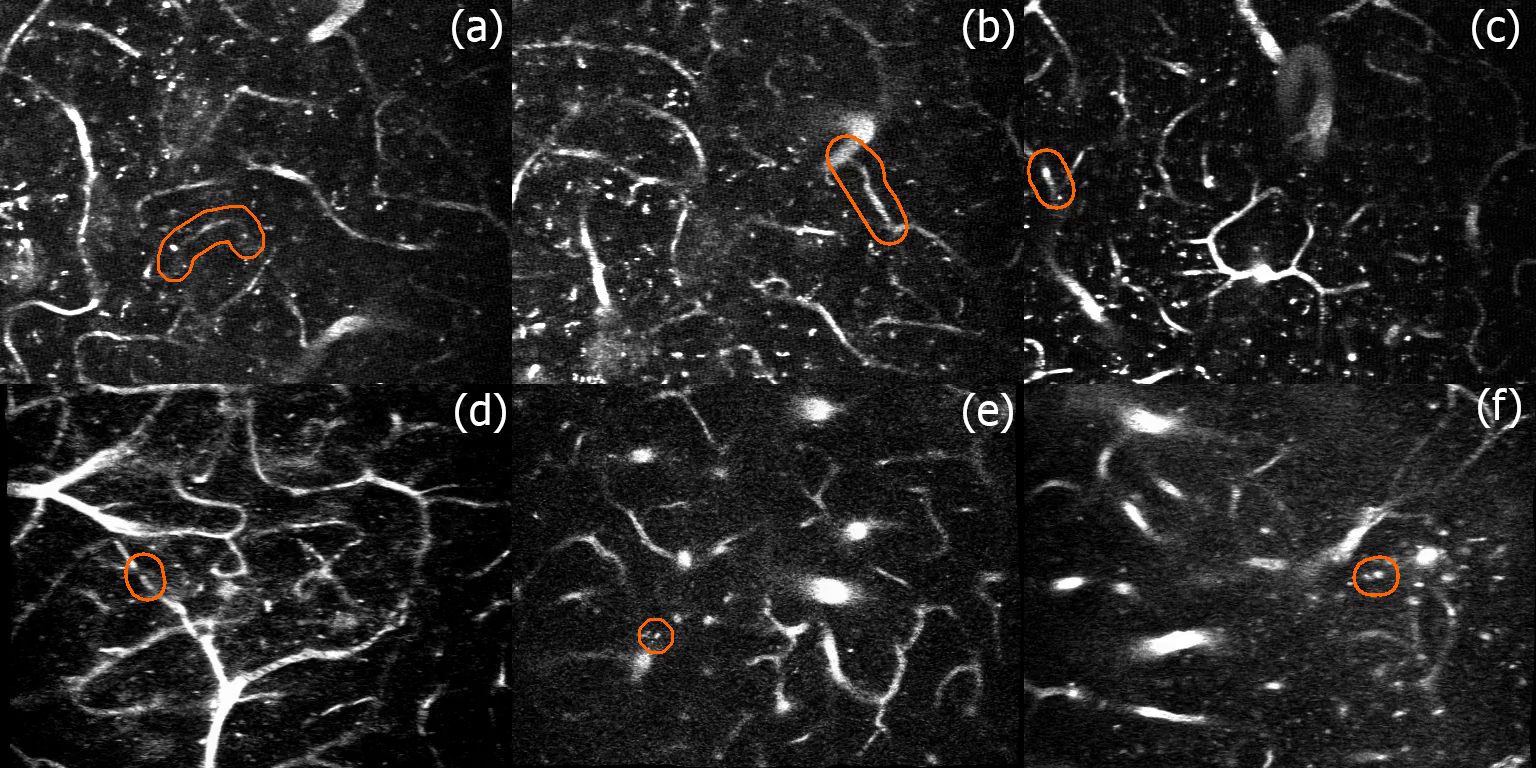}
 \caption{Samples of video frames from imaged mouse brain tissue with annotated vessel contours (orange) for (a, b, c) flowing and (d, e, f) stalled vessels.\label{fig1}}
\end{figure*}

The $z$ axis of 3D image volumes represents both depth (looking at successive layers of brain tissue) and time, such that a step downward in $z$ is also a step forward in time. We refer the reader to the Stall Catcher's Tutorial \cite{youtube2018tutorial} for further details of image acquisition. 3D image volumes were converted to mp4 video files. The dataset contained over $580,000$ videos of blood vessels in brain tissue with a total size of $1.5$ terabytes. The target vessel segment for each set of images was annotated with a contour. Figure~\ref{fig1} shows example images with annotated vessels.

The dataset also contained metadata for each video, including identifier for the research project that generated the video (\texttt{project-id}), crowd-labeled probability that the vessel is stalled, ranging between 0 for \textit{flowing} and 1 for \textit{stalled} (\texttt{crowd-score}), and a Boolean variable indicating a highly confident label (\texttt{tier1}). Additionally, a subset of videos has been reviewed by an expert, who has labeled videos as either \textit{stalled} or \textit{flowing}. When an expert label is not available, \textit{stalled} vessels were defined as videos with crowd scores greater than or equal to 0.5, while videos with crowd scores less than 0.5 are labeled as \textit{flowing}. The majority of crowd scores gave confident labels with the values close to either 0 or 1. Therefore, a threshold of 0.5 for making binary labels was a reasonable option for these data~\cite{bracko2020high}.

The \texttt{tier1} flag column indicates the highest quality data. These are videos that either have an expert label or a highly confident crowd label. A highly confident crowd label was defined as one with a crowd score equal to 0 (surely flowing) or greater than or equal to 0.75 (most likely stalled).
Thus, videos with \texttt{tier1} flag can be seen as the most reliable examples of stalled or flowing vessels.

The dataset was highly unbalanced, with 99.7\% of videos displaying healthy vessels.
There were 1,887 videos with stalled vessels, with only 706 of them had a high confidence label (\texttt{tier1=1}).
The rest 57,1161 videos contained flowing vessels.
To facilitate easier model prototyping, organizers provided a "micro" subset of the whole dataset, with the 70/30 ratio of stalled to healthy vessels.
For our experiments, we used all 1,887 videos of stalled vessels from the "micro" subset and 50K videos of flowing vessels sampled both from the "micro" subset and from the full dataset.
To alleviate class imbalance, we employed heavy data augmentations and undersampling of the \textit{flowing} class as described below.
The majority of videos had $512\times384$ resolution, with some samples with $418\times384$ resolution.
The distribution of data in "micro" train and test subsets in each resolution is shown in Table~\ref{tab1}.
Finally, the number of frames in videos varied between 20 and 300, with 60 frames on average.

\begin{table}[htb]
 \caption{Distribution of video resolution formats in "micro" train and test data sets. \label{tab1}}
 \begin{centering}
 \begin{tabular}{ccc}
 \hline
 & \multicolumn{2}{c}{\textbf{Resolution}}\\
 \hline
 \textbf{Image set}	& \textbf{$512\times384$}	& \textbf{$418\times384$}\\
 \hline
 Train		& 2380			& 19 \\
 Test		& 14037			& 123\\
 \hline
 \end{tabular}
 \end{centering}
\end{table}

\subsection{Data preprocessing and augmentations}
The region of interest (ROI) containing the vessel segment was annotated in all videos with the orange contour (see Figure~\ref{fig1}).
We trained models both on whole video frames and on cropped ROIs.
For the latter, we cropped selected 3D rectangular cuboid around the ROI in all videos.
The ROI crop size was usually smaller than the video frame size.
This helped models to focus on relevant information, as well as reduce training time and required computational resources.

\begin{figure}[htb]
 \centering
 \includegraphics[width=0.85\linewidth]{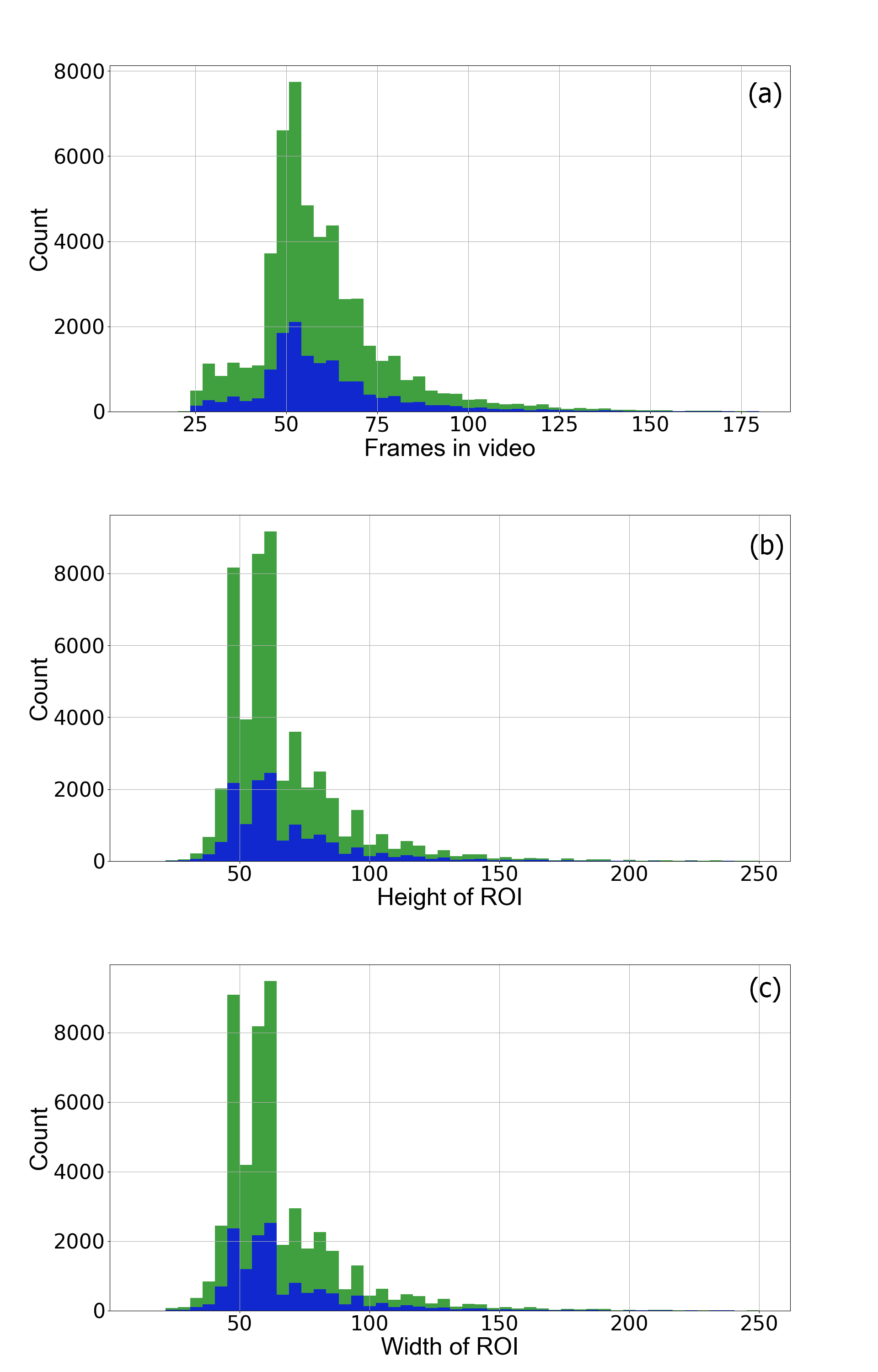}
 \caption{Distribution of ROI sizes for train (green) and test (blue) sets.\label{fig2}}
\end{figure}

Figure~\ref{fig2} shows the distribution of ROI sizes for training and testing sets. The ROIs were around $60 \times 66 \times 67$ pixels, so it was possible to fit several 3D samples into a single GPU (NVIDIA 1080Ti, 11 GB).
However, all 3D samples had different sizes.

Since the dataset was highly unbalanced, intensive data augmentations were a crucial part of our training pipeline.
None of the existing data augmentation libraries satisfied our requirements for 3D image data transforms.
Therefore, we created a custom library dubbed Volumentations~\cite{volumentations}, inspired by the open-source library Albumentations~\cite{albu}, that is widely used for 2D images augmentations.

For augmenting 3D samples during training, we used either random mirroring of the 3D image along each of three axes ("mirror 3 axes") or the following complex transformation protocol ("heavy augs"), where $p$ shows the probability of use:

    \textbf{Spatial transformations:}
    \begin{itemize}
    \item Random rotation up to 10 degrees along the first axis, $p=0.3$
	\item Elastic transform with interpolation, $p=0.1$
    \item Rotation by 90 degrees
    \item Flip along each of the axes, $p=0.5$
    \item Grid dropout, $p=0.1$
    \end{itemize}

    \textbf{Pixel transformations:}
    \begin{itemize}
    \item Adding Gaussian noise, $p=0.2$
    \item Random Gamma blurring, $p=0.2$
    \end{itemize}

    \textbf{Custom-made transformations:}
    \begin{itemize}
     \item Random crop from borders, $p=0.4$
     \item Random drop plane, $p=0.5$
     \item Resize, $p=1.0$
    \end{itemize}

Some of used 3D transforms were adapted from their 2D counterparts implemented in Albumentations~\cite{albu}, including \texttt{Rotate}, \texttt{ElasticTransform}, \texttt{GridDropout}, \texttt{Flip}, \texttt{RandomRotate90}, \texttt{GaussianNoise}, \texttt{RandomGamma}.
Other augmentations were developed specifically for this task. For example, \texttt{RandomCropFromBorders} randomly deletes some pixels from one of the 3D sample borders, while \texttt{RandomDropPlane} randomly deletes intermediate 2D planes from some 3D sample axes.
The use of Volumentations and the "heavy augs" protocol significantly improved the classification performance (see Results).

\subsection{Models}
\begin{figure}[htb]
 \includegraphics[width=0.99\linewidth]{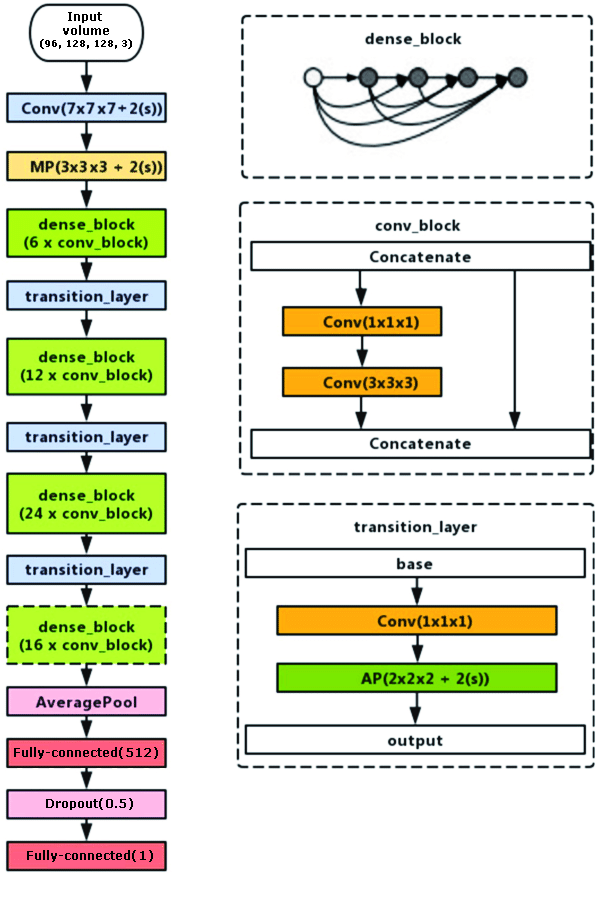}
 \caption{Architecture of 3D DenseNet121. Input dimensions correspond to $frames\times height\times width\times channels$.\label{fig3}}
\end{figure}

We used 3D CNNs based on \texttt{Conv3D} layers for this classification task. We re-designed multiple commonly-used 2D CNNs architectures (ResNet~\cite{resnet}, ResNeXt~\cite{xie2017aggregated}, SE-ResNeXt~\cite{Hu_2018_CVPR}, and DenseNet~\cite{densenet}) for working with 3D data by replacing all 2D convolutions by \texttt{Conv3D} layers, and changing all other layers respectively, but keeping the principle architecture and the number of layers.
The architecture of 3D DenseNet121 is shown in Figure~\ref{fig3}.
This model has a relatively low number of parameters (11.9M), can accommodate the larger batch size, and was the best for this task according to classification results (see Results).

\begin{figure}[htb]
 \includegraphics[width=0.99\linewidth]{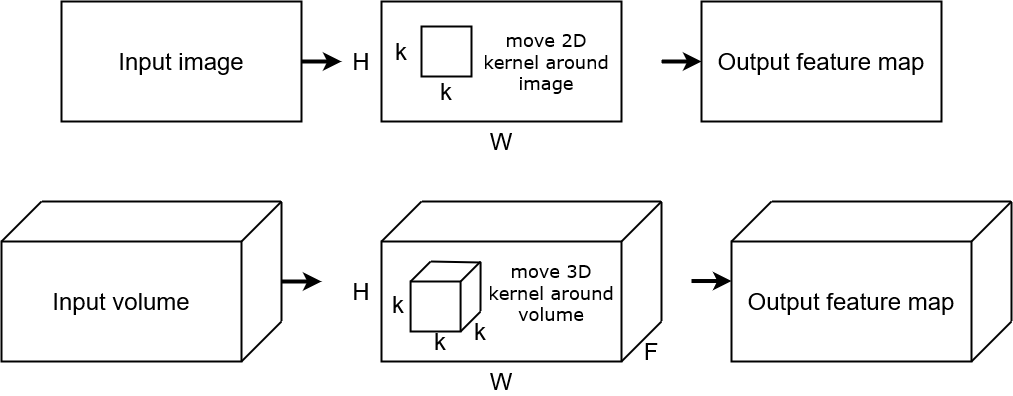}
 \caption{Schematic representation of 2D and 3D input samples, convolutions, and output feature maps.
 The 2D sample is an image with dimensions $H\times W$, where $H$ and $W$ are the height and width of the image. The 2D convolutions are 2-dimensional matrices with weights (kernels) with the $k\times k$ size; in our case, $k = 3$.
 The 3D sample shape is $(F, H, W)$, where $F$, $H$ and $W$ are frames, height and width of the video sample, respectively. The 3D convolutions are moved across the volume of the sample and are 3-dimensional matrices with the size of $ k\times k \times k, k = 3$.}
 \label{fig2D3D}
\end{figure}

2D models pre-trained on the ImageNet~\cite{imagenet} dataset have previously demonstrated great potential for transfer learning into different domains with increased accuracy and/or faster convergence~\cite{ternausnet}, including applications to medical image analysis~\cite{rakhlin2019breast,solovyev2020bayesian,kalinin2020medical,rsna}.
To leverage pre-training on large-scale 2D image datasets for 3D image analysis, we developed a method for weight transfer from 2D to 3D CNNs.
Below, we illustrate the weights transfer by a simple concrete example with a 3D video.
Figure~{\ref{fig2D3D}} shows the diagram of 2D and 3D input samples, convolutions, and output feature maps.

For instance, a 3D sample input has following dimensions: $(F, H, W, C)$, where $F$, $H$ and $W$, are number of frames, height and width, and $C$ is number of channels. Let's assume $C=1$ for simplicity (such as in gray scale images), then 3D sample shape is $(F, H, W, 1)$ = $(F, H, W)$.
Now, let's take a random frame $K$ (for example, $K=2$): it will be an image with dimensions $H\times W$.
We can compute a feature map for this frame $K$ using \texttt{Conv2D}.
Let's take a point with $i, j$ coordinates and consider $3\times3$ region in the vicinity of this point, assuming the input frame has the following values around this point:
\begin{equation*}
F =
\begin{pmatrix}
4 & 3 & 2\\
1 & -1 & 6\\
7 & 1 & 0
\end{pmatrix}
\end{equation*}

Consider a \texttt{Conv2D} convolution with kernel size $3\times3$ and weights:
\begin{equation*}
W = \left(
\begin{array}{cccc}
W_{11} & W_{12} & W_{13}\\
W_{21} & W_{22} & W_{23}\\
W_{31} & W_{32} & W_{33}
\end{array}
\right)
\end{equation*},

For example, if we take a \texttt{Conv2D} convolution with weights of:
\begin{equation*}
W = \left(
\begin{array}{cccc}
2 & 3 & 4\\
-3 & 0 & 1\\
2 & 3 & 6
\end{array}
\right)
\end{equation*}
the output at the feature map for $[i,j]$ point is:
\begin{multline}
Out[i, j] = 2\cdot4 + 3\cdot3 + 4\cdot2 + (-3)\cdot1 + \\
0\cdot(-1) + 1\cdot6 + 2\cdot7 + 3\cdot1 + 6\cdot0 = 46
\end{multline}

Now, let's consider the 3D volume around the same point in the vicinity $3\times3\times3$. We now take 3 frames around the point: one frame before and one after the previously considered frame. For illustration, we assume they have the following values:
\begin{equation*}
F1 =
\begin{pmatrix}
3 & 4 & 1\\
1 & 0 & 6\\
7 & 1 & 0
\end{pmatrix}
F2 =
\begin{pmatrix}
4 & 3 & 2\\
1 & -1 & 6\\
7 & 1 & 0
\end{pmatrix}
F3 =
\begin{pmatrix}
5 & 3 & 2\\
1 & 0 & 7\\
7 & 2 & 1
\end{pmatrix}
\end{equation*}

We can define a 3D convolution \texttt{Conv3D} with kernel $3\times3\times3$ in the following way:
\begin{equation}
W1 = W3 =
\begin{pmatrix}
0 & 0 & 0\\
0 & 0 & 0\\
0 & 0 & 0
\end{pmatrix}
W2 =
\begin{pmatrix}
W_{11} & W_{12} & W_{13}\\
W_{21} & W_{22} & W_{23}\\
W_{31} & W_{32} & W_{33}
\end{pmatrix}
\end{equation}

Here, if we take $W2$ the same as our \texttt{Conv2D} convolution for the second frame, and other parts $W1$ and $W3$ as zero matrices, at 3D point $[2, i, j]$ we will get the same output as in the 2D case:
\begin{multline}
Out[2, i, j] = 0\cdot3 + 0\cdot4 + 0\cdot1 + ... +\\
     + 2\cdot4 + 3\cdot3 + 4\cdot2 + ... + 2\cdot7 + 3\cdot1 + 6\cdot0 + \\
    0\cdot5 + 0\cdot3 + 0\cdot2 + ... + 0\cdot1 = 46
\end{multline}

Applying similar $3\times3\times3$ kernels to all frames in the input volume, we get a set of feature maps for each frame of the 3D input, in the same manner as if we computed feature maps separately for each frame using \texttt{Conv2D}.
Such initialization of weights can be transferred from 2D CNNs pre-trained on ImageNet (or other datasets) to 3D networks of the same architecture. It provides reasonable initialization as intermediate feature maps contain useful features, even though they are applied to some video frames.

Developing this idea further, we can assume that neighboring frames differ only slightly from each other.
Then, we can average feature maps for the close frames.
In order to do so, we create a 3D kernel $3\times3\times3$ with weights re-distributed among the neighboring frames, as:
\begin{equation}
W1 = W2 = W3 =
\begin{pmatrix}
W_{11}/3 & W_{12}/3 & W_{13}/3\\
W_{21}/3 & W_{22}/3 & W_{23}/3\\
W_{31}/3 & W_{32}/3 & W_{33}/3
\end{pmatrix}
\end{equation}

For example,
\begin{equation}
W1 = W2 = W3 =
\begin{pmatrix}
2/3 & 3/3 & 4/3\\
-3/3 & 0/3 & 1/3\\
2/3 & 3/3 & 6/3
\end{pmatrix}
\end{equation}

\begin{table*}[htb]
\centering
\caption{Common model hyper-parameters.}
  \begin{tabular}{lc}
  \toprule
  Parameter & Description \\
  \midrule
      Optimizer & AdamAccumulate (accumulate for 20 iterations)  \\
      Initial learning rate & 1e-4  \\
      Learning rate scheduler& ReduceLROnPlateu with Decay \\
      Decay& 0.95 every 3 epochs without improvements \\
      Loss function & Binary cross-entropy  \\
  \bottomrule
  \end{tabular}
  \label{table:params}
\end{table*}

With such weight re-distribution, we now take into account frames $F1$ and $F2$ while keeping the scale of the values in the output feature map. For the above example, the output at point $[2, i, j]$ will be:
\begin{multline}
OUT[2, i, j] = (2\cdot3 + 3\cdot4 + 4\cdot1 + ... +\\
    + 3\cdot2 + 6\cdot1) / 3 = (42 + 46 + 57)/3 = 48.33
\end{multline}

Using this technique, we converted weights from 2D CNNs pre-trained on the ImageNet dataset to initialize their 3D counterparts.

\subsection{Evaluation}
The challenge submissions were evaluated using Matthews correlation coefficient (MCC), defined as follows \cite{matthews1975comparison}:
\begin{equation}
\text{MCC}=\frac {TP\times TN-FP\times FN}{\sqrt {(TP+FP)(TP+FN)(TN+FP)(TN+FN)}}
\end{equation}
where $TP$ – true positives, $FP$ – false positives, $TN$ – true negatives, and $FN$ – false negatives.
This metric takes into account all four components of the confusion matrix and is well-suited for unbalanced datasets.

To compare different models, we computed the area under the receiver operating characteristic curve (AUC) from their predictions. The receiver operating characteristic curve plots the True Positive Rate (TPR) versus False Positive Rate (FPR) at different classification thresholds. The TPR and FPR are defined as follows:
\begin{equation}
\text{TPR}=\frac {TP}{TP+FN}
\end{equation}
\begin{equation}
\text{FPR}=\frac {FP}{FP+TN}
\end{equation}

The AUC measures the entire two-dimensional area underneath the receiver operating characteristic curve.

The models were evaluated using the AUC and MCC metrics and five-fold cross-validation (CV) scheme.
The training set was divided into five parts, stratified by the stalled vessels, with 80\% of samples used for training and 20\% for validation.
The $TP$, $FP$, $TN$, and $FN$ were calculated using all validation samples in each of five folds, thus providing validation metrics for a full dataset.
The holdout testing set was predicted with each of these five-fold models.
The predictions from five models were averaged and used to calculate the final metrics on the testing set.

\subsection{Model training}

Each model was trained on the total of 51,490 videos (49,603 "non stalled", 1,887 "stalled"). Stalled videos included \texttt{tier1} as well is non-\texttt{tier1} videos.
We used a range of 3D DenseNet and ResNet models with weights transferred from their 2D counterparts pre-trained on the ImageNet dataset.
The models were trained using five fold cross-validation scheme.
The hyper-parameters used for the training experiments are outlined in Table 2.

Here are the details of the training process for the best single model 3D DenseNet121 (public leaderboard MCC score: 0.8436, public leaderboard MCC score: 0.8411).

\begin{itemize}
\item As the training dataset was highly unbalanced (see Section 2), we used a sampler as an extra balancing technique. Batches were randomly generated with 25\% of class 1 and 75\% of class 0.

\item Augmentations: heavy augmentations were applied (see Section 4).

\item Additional fully connected layer was added after backbone before classification layer; we used dropout of 0.5 to prevent overfitting.

\item The training process included two stages. At the first stage, the MCC metric was maximized on validation. Then, starting from the weights of the first stage, we optimized the AUC metric.

\item At the first stage, validation was performed only on the part of available data, such that class "0" and class "1" were 50/50\% balanced. On the second run, we used all available validation data with class "0" comprising 80\% and class "1" - 20\%.
\end{itemize}

\subsection{Model interpretability}
It can be useful to visualize the relevance of pixels in the input image to the classification of that image to make sure the prediction relies on blood flow and not on irrelevant information such as image artifacts.
To visualize model attention, we used global max pooling instead of average pooling before the classification head.
It facilitated highlighting regions that formed feature vectors, as they were determined by the maximum pixel in each feature map.

For better clarity of heatmaps, it is critical to use higher input resolution. For instance, for input with $N \times H \times W$ size ($N$ – number of frames, $H$ - height, and $W$ - width), corresponding heatmap size is $N/32 \times H/32 \times W/32$. So, for the input 3D sample with size $128 \times 160 \times 160$ the heatmap size is only $4 \times 5 \times 5$ pixels.

We used inputs with dimensions $320 \times 400 \times 400$ px for creating heatmaps with size $10 \times 13 \times 13$ px. We scaled the heatmap 32 times to match the input resolution and overlaid them with inputs.
We used the last output from the convolution block before the \texttt{Global Max Pooling} layer to plot heatmaps. In our case, feature maps are feature volumes. For 3D ResNet18, their dimensions were (10, 13, 13, 512), i.e., 512 feature volumes with shapes (10, 13, 13) are transformed into a 1D vector of 512 values after 3D global max pooling.

To visualize heatmaps, we calculated maximum, mean, and standard deviation for each point in the corresponding $(10, 13, 13)$ volume. Using vectorization in NumPy~\cite{harris2020array}, they can be computed as following:
\begin{python}
# Layer before pooling
pstd = np.std(out, axis=-1)
pmax = np.max(out, axis=-1)
pmean = np.mean(out, axis=-1)
# Normalize to 0..255 to plot as RGB
nstd = normalize-array(pstd, 0, 255)
nmax = normalize-array(pmax, 0, 255)
nmean = normalize-array(pmean, 0, 255)
# Stack as RGB video
hmap = np.stack((nstd, nmax, nmean), axis=-1)
\end{python}

We used maximum, mean, and standard deviation instead of RGB channels.
Maximum shows the maximum value possible in this point, standard deviation shows how the weight varies at this point, and mean corresponds to the average influence of this point for the final classification.

\begin{figure}[htb]
 \includegraphics[width=0.99\linewidth]{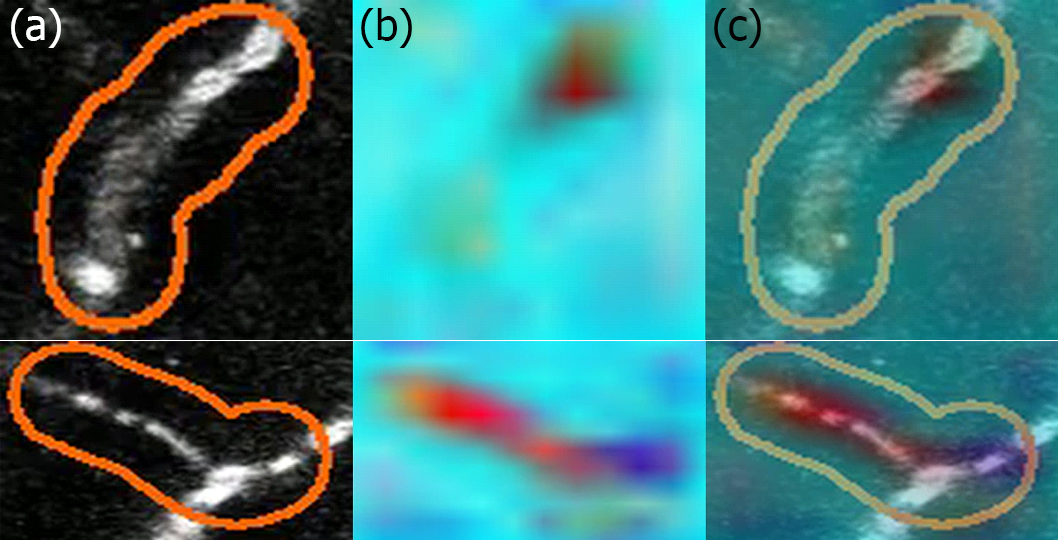}
 \caption{Model attention visualization: (a) input slices cropped around ROI, (b) scaled heatmaps, and (c) overlays of heatmaps with inputs.\label{fig:heatmap}}
\end{figure}

\section{Results}
\label{results}
\subsection{Ablation study}
First, we illustrated the influence of augmentations on the classification performance of the DenseNet121 model. As shown in Table~\ref{tab2}, using augmentations led to the higher AUC and MCC metrics.

\begin{table*}[htb]
\centering
 \caption{Classification metrics achieved with and without augmentations on a validation set. \label{tab2}}
 \begin{tabular}{cccccc}
\toprule
 Model	& Input size	& Augmentations & AUC &  MCC \\
\midrule
 DenseNet121	& (96, 128, 128, 3) & no augs &0.9073&  0.5622\\
 DenseNet121    & (96, 128, 128, 3) & heavy augs & 0.9665 & 0.7226\\
\bottomrule
\end{tabular}
\end{table*}

We also assessed the utility of 2D-to-3D pre-trained weight conversion for transfer learning.
Our experiments showed that weight transfer from 2D model pre-trained on the ImageNet dataset to 3D model robustly helped to decrease convergence time and to improve classification performance results.
Specifically, we achieved the MCC of 0.810 with DenseNet121 initialized by 2D weight transfer, compared to 0.795 with random initialization.
The models with converted weights are available at~\cite{classmodels3Dzoo}.

\subsection{Vessel classification results}
First, we trained 3D models on the whole video frames rescaled into different sizes, such that they could fit into the 3D model. The results for different input sizes and model backbones are shown in Table~\ref{tabm1}. The input dimensions of the samples here correspond to $frames \times height \times width \times channels$.
Best performance metrics were reached using a larger input resolution of $192\times256$, which suggested that smaller resolutions were not enough to capture all important information. This motivated us to perform further experiments using ROI cropping to focus on relevant areas in the image, while reducing memory footprint.
\begin{table*}[htb]
\centering
 \caption{Results for models trained on videos without ROI cropping. \label{tabm1}}
 \begin{tabular}{lccccc}
\toprule
Model	& Input size	& Batch size & MCC, Valid & AUC, Valid & MCC, Test \\
\midrule
 ResNet18		& (64, 96, 128, 3) & 20	& 0.534  & 0.821 & 0.4437\\
 ResNet18       & (64, 192, 256, 3)& 6 & 0.669 &  0.892 & 0.6105\\
 DenseNet121	& (64, 96, 128, 3) & 10 &0.534 &  0.833 &  0.4201\\
 ResNet34       & (128, 96, 128, 3)& 5  &0.544 & 0.840 & 0.4251\\
 ResNet50       & (64, 192, 256, 3)& 4 & 0.697 & 0.914 & 0. 5986\\
\bottomrule
 \end{tabular}
\end{table*}

\begin{table*}[htb]
\centering
 \caption{Results for models trained on cropped ROI regions. \label{tabm2}}
 \begin{tabular}{lccccc}
\toprule
 Model	& Input size	& Parameters & MCC, Valid & AUC, Valid & MCC, Test \\
\midrule
 ResNet18		& (96, 128, 160, 3) & Mirror 3 axes &  0.720 & 0.927& 0.6734\\
 ResNet18       & (128, 160, 160, 3)& Heavy augs, Sampler 25\%/75\%& 0.623 & 0.966 & 0.7234\\
 DenseNet121	& (96, 128, 160, 3) & Mirror 3 axes &0.837 & 0.959& 0.7442\\
 DenseNet121    & (96, 128, 128, 3) & Heavy augs & 0.641& 0.967& 0. 7887\\
 DenseNet121    & (96, 128, 128, 3) & Heavy augs, Sampler 25\%/75\% & 0.649& \textbf{0.973}& 0.8129\\
 DenseNet121    & (96, 128, 128, 3) & tier1 only, extra-heavy augs, Sampler& 0.622 & 0.958 & \textbf{0.8436}\\
 DenseNet169    & (64, 128, 128, 3)& Heavy augs & &  & 0.7485\\
 DenseNet201    & (96, 128, 128, 3)& Heavy augs, Sampler 25\% /75\% & 0.588& 0.963& 0.7532\\
 Best ensemble    & --- & --- & --- &  --- & 0.8555\\
\bottomrule
 \end{tabular}
\end{table*}

ROI cropping allowed us to use images of higher resolution without requiring more training time or computational resources.
On the cropped regions, we tested multiple 3D CNN architectures with different backbones and hyper-parameters, as shown in Table~\ref{tabm2}.
The 3D DenseNet121 demonstrated the overall best classification performance for this task, as measured by AUC and MCC.
This model has a relatively low number of parameters (11.9M) and can accommodate a larger batch size, which reduces training time.
These results stress the importance of using heavy augmentations and image sampling to correct for class imbalance.

To highlight the areas in the image that influenced the network's decisions, we created model attention heatmaps shown in Figure~\ref{fig:heatmap}.
These results indicate that the last layer of the 3D CNN focused on the blood vessel in general, since the shape of the active region in the heatmap overlaps well with the shape of the vessel.
By compiling these heatmaps, we created a video visualization of many analyzed vessels that is available on YouTube~\cite{youtube}.

\subsection{Model reliability assessment}
\subsubsection{Probability calibration}

When model predictions can influence important decisions such as a medical diagnosis, it is important to assess whether a prediction can be trusted or not. A standard approach to doing so is to use the classifier\textquotesingle~s own confidence, i.e., probabilities from the softmax layer of a neural network. While using model\textquotesingle s own implied confidences appears reasonable, it has been shown that the raw confidence values from modern neural networks are often poorly calibrated~\cite{guo2017calibration}. For example, neural networks with residual connections~\cite{resnet} typically exhibit average confidence that is substantially higher than their accuracy~\cite{guo2017calibration}.

\begin{figure}[htb]
 \includegraphics[width=0.99\linewidth]{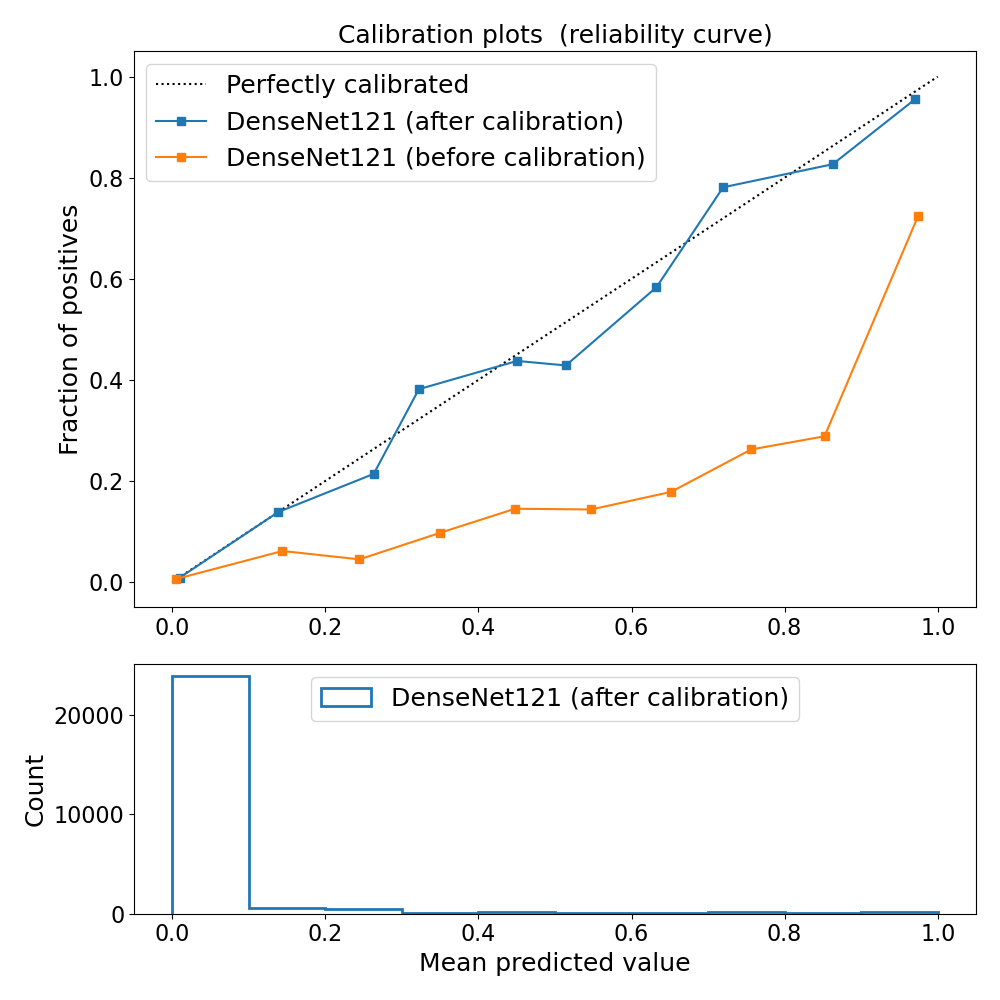}
 \caption{Top: performance of the 3D DenseNet121 models before (orange) and after (blue) calibration. Bottom: histogram of the mean predicted values after calibration.
 \label{fig_calibration}}
\end{figure}

We assessed our best model's trustworthiness by visualizing model calibration as a reliability diagram in Figure~\ref{fig_calibration}.
The model is clearly miscalibrated as it consistently outputs probabilities higher than its accuracy.
To correct the model's overconfidence, we performed probability calibration based on isotonic regression~\cite{calibration}, arguably the most common non-parametric calibration method~\cite{guo2017calibration} that learns a piecewise constant function to transform uncalibrated outputs.
Since we didn't have access to the labels of the test set, we randomly split the validation set into two equal parts.
We used the first half of the validation set to train a calibration classifier and predicted calibrated probabilities on the second half. As shown in Figure~\ref{fig_calibration}, this approach was effective at calibrating predictions, improving the trustworthiness of our model.
The bottom panel shows the histogram of the mean predicted values after calibration.
Unsurprisingly, most of the predicted probabilities are close to zero due to the highly unbalanced dataset.

\subsubsection{Uncertainty estimation}
Another approach to assess the reliability of network decisions is to obtain an estimate of Bayesian uncertainty.
While mathematically sound tools to assess model uncertainty in the Bayesian probability theory framework exist, they are typically very expensive computationally.
However, it has been shown that the use of dropout~\cite{srivastava2014dropout} in neural networks can be interpreted as a Bayesian approximation of a Gaussian process that provides an estimate of the probability distribution over many possible models for a given sample~\cite{gal2016dropout}.
Therefore, dropout can be used as a tool to represent uncertainty in deep learning models, including deep convolutional neural networks.

\begin{figure}[htb]
 \includegraphics[width=0.99\linewidth]{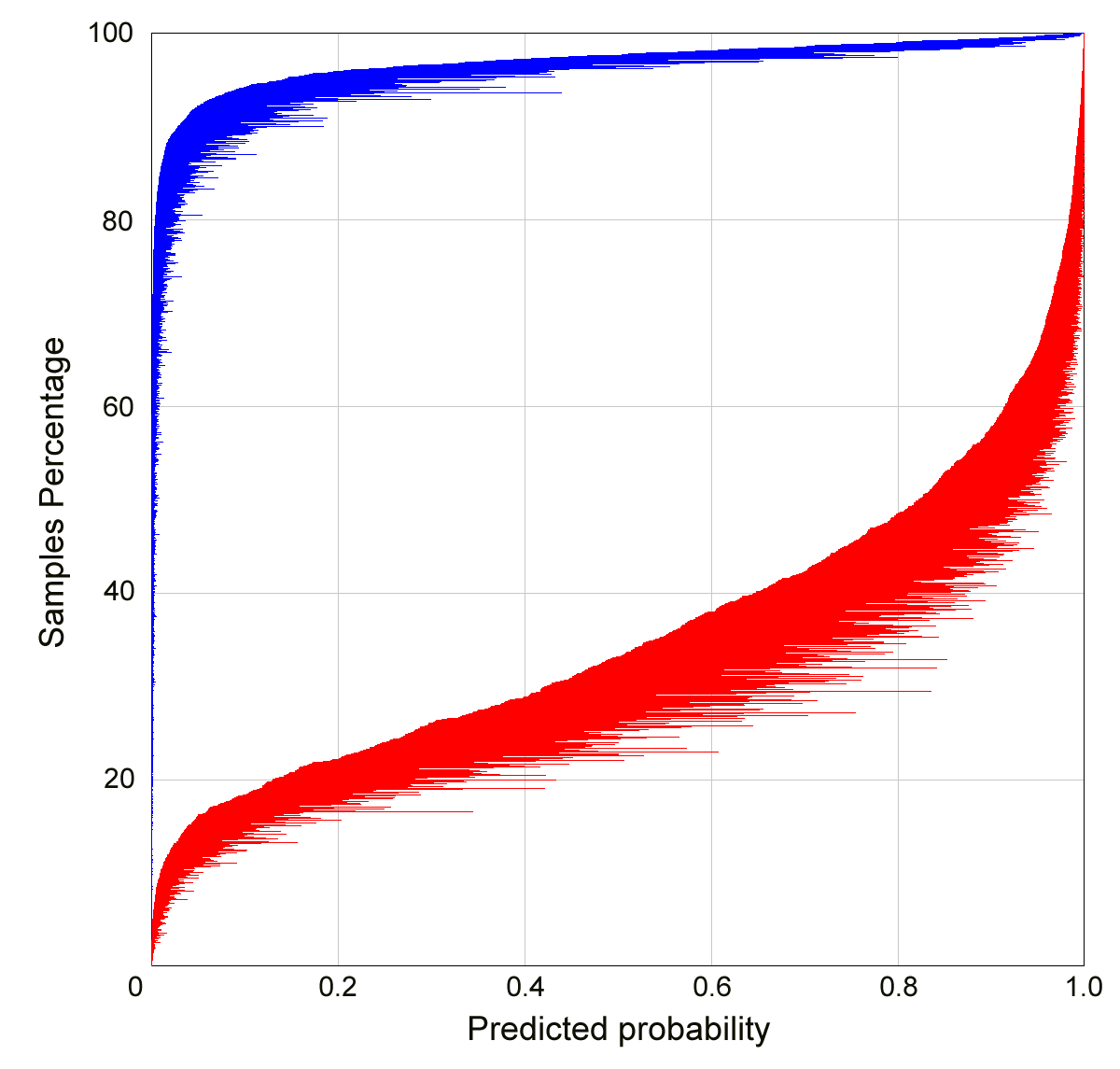}
 \caption{The probabilities distribution for the flowing (blue) and stalled (red) samples calculated 50 times for the 5 folds of the 3D  DenseNet121 models with the active dropout of 0.5.\label{fig_uncert}}
\end{figure}

To assess the reliability of decisions on these data, we computed the uncertainty of our best-performing model using dropout.
We randomly selected 5000 flowing samples (target label 0) and all 1887 stalled samples (target label 1).
For each sample, we calculated the predicted probabilities 50 times for five folds using the 3D DenseNet121 model with the active dropout of 0.5~\cite{gal2016dropout}.
The results are presented in Figure~\ref{fig_uncert}.
Here, the blue lines show the output probability variations for the flowing samples and red shows the results for the stalled samples.
Overall, the model has higher variation in the output probabilities (has more uncertainty) for the stalled samples, which was expected, given high class unbalance in the data.
The flowing samples (blue) had a mean probability difference of $0.0229$ and a standard deviation of $0.0518$. While the stalled samples (red) had a mean difference of $0.0924$ with a standard deviation of $0.0922$.

Since one of the goals of this competition was to improve the automatic labeling of images, this procedure can help with better identification and processing of uncertain inputs that may require attention of an expert to produce the correct label.

\section{Discussion}
Common approaches to video classification are typically based on extracting features from the sequence of the video frames using 2D CNNs and then using those features as inputs to LSTM models~\cite{islam2020review}. However, these methods did not demonstrate competitive performance for this task. Other approaches that performed well in the challenge also employed 3D CNNs. The second-place winner used 3D ResNets and cropped regions of interest on videos, similarly to our solution. The third-place winner used spatio-temporal neural network architectures based on ResNet18. They were designed for video classification and pre-trained for action recognition on the Kinetics-400 data set~\cite{lipstein2020winners}.

To assess the potential usefulness of our model for automatic labeling, it is essential to compare model predictions with the human labeling approach.
Currently, specialized methods aggregate multiple players into a single "crowd answer" that achieves a 0.99 sensitivity and 0.95 specificity threshold on 95\% of datasets.
Our model gives higher specificity of 0.99 at the cost of the lower sensitivity of 0.85.
The precision-recall trade-off can be adjusted to fit research and clinical needs by choosing the threshold for binary classification.
The results indicate a potential for multiple possible ways of combining machine learning produced annotations and human labeling, which can flexibly address current limitations of data pre-processing and analysis and provide insights towards mechanistic connections between stalled blood flow and AD.

We should note that multiphoton microscopic imaging is an invasive diagnostic technique. This imaging method can see down hundreds of micrometers to a millimeter into brain tissue, but only after gaining optical access to the brain surface by removing a section of the skull~\cite{hernandez2019neutrophil,haft2019deep}.
It is not likely that such an approach would be used in any human clinical diagnostic. However, the study of
animals could bring a better understanding of the mechanisms in brain vasculature dysfunctions such as stalled blood flow, which in turn can help to identify mechanisms behind AD and manage it.

\section{Future improvements}
The challenge organizers provided outlines for vessels in each video, produced by their segmentation model~\cite{haft2019deep}, such that we only had to analyze already pre-selected areas by cropping the ROI around them.
To automate the solution even further, one could consider performing simultaneous classification and segmentation of vessels, which could potentially alleviate the requirements of labeling images twice---for localization and for classification.
Heatmaps produced by our approach outline vessels of interest well and could be employed as weak labels in the localization or segmentation problem setting.

It is worth noting that the difference between two consecutive frames of the video samples was relatively small. This leads to the idea of considering only every n-th frame as an input of the model and potentially achieving similar performance at a fraction of computational time. Such investigation is out of the scope of this paper, however, we agree with the suggestion of one of the reviewers that this could help reducing model's computational requirements.

Another possibility for improvement is using the crowd score itself. The crowd score values were also provided, which allowed experimenting with the threshold values for creating binary target labels. The majority of the crowd scores gave confident labels with values close to either 0 or 1. One can add another head to the models with a regression task to predict crowd score values. A similar multi-tasking strategy (regression plus classification) demonstrated model improvements in the other computer vision tasks \cite{rsna}.

\section{Conclusion}
In this paper, we described a solution for automatic brain vessel classification in a highly unbalanced dataset of 3D images.
This approach was used to produce a winning solution to the "Clog Loss: Advance Alzheimer’s Research with Stall Catchers" machine learning competition, which posed a binary classification problem of identifying blood vessels in live mouse brain as \textit{flowing} or \textit{stalled}.
Our best solution was based on 3D CNN architecture trained on pre-selected crops with custom 3D image augmentations, balanced sampling, and test-time augmentations.

The ablation study has shown the significant influence of data augmentation on the performance of the model. We made a custom library with 3D data augmentations, Volumentations~\cite{volumentations}, publicly available.

We tested a range of 3D classification models with weights transferred from their 2D analogs pre-trained on the ImageNet dataset. 3D DenseNet121~\cite{huang2017densely} demonstrated the best performance for this task, achieving the MCC of 0.84 and AUC of 0.97 for a single model, which by itself was enough to take the first place in the competition.

Our final submission to the competition included an ensemble of 18 models with different backbones, demonstrating even further improvement of the MCC to 0.855.
The source code for our solution is publicly available at GitHub~\cite{dd2020solution1stplace}.

\bibliographystyle{cas-model2-names}

\bibliography{main}

\end{document}